# Superradiance of a 2D-spaser array

A.V. Dorofeenko,[1,2] A.A. Zyablovsky,[1,2] A.P. Vinogradov,[1,2] E.A. Andrianov,[1,2] A.A. Pukhov,[1,2]
A.A. Lisyansky[3]*

[1]Institute for Theoretical and Applied Electromagnetics RAS, Moscow, Russia

[2]Moscow Institute of Physics and Technology, Dolgoprudny, Moscow reg., Russia

[3]Department of Physics, Queens College of the City University of New York, Flushing, NY 11367, USA

We demonstrate that interacting spasers arranged in a 2D array of arbitrary size can be mutually synchronized allowing them to supperradiate. For arrays smaller than the free space wavelength, the total radiated power is proportional to the square of the number $N$ of spasers. For larger arrays, the radiation power is linear in $N$. However, the emitted beam becomes highly directional with intensity of radiation proportional to $N^2$ in the direction perpendicular to the plane of the array. Thus, spasers, which mainly amplify near fields, become an efficient source of far field radiation when they are arranged into an array.

*alexander.lisyansky@qc.cuny.edu

Recent developments of nanotechnologies incorporating plasmonic structures have led to the design of new generation of components for optoelectronics and devices operating in the deep subwavelength regime [1-3]. In particular, new light sources – spasers (surface plasmon amplification by stimulated emission of radiation) – have been proposed [4] and experimentally demonstrated [5]. Schematically, the spaser is an inversely populated two-level system (TLS), e.g. an atom, a molecule, or a quantum dot, interacting with a plasmonic nanoparticle (NP) [4, 6] or with a plasmonic waveguide via near field [7-9]. The transition from the excited to the ground state is accompanied by oscillations of the TLS dipole moment. These oscillations excite surface plasmons at the NP. Due to the short distance between the NP and the TLS, plasmon generation is much more efficient than photon radiation. In turn, plasmon oscillations induce the TLS to radiate providing feedback for the spaser.

The main sources of losses in spasers are dissipation in the metal NP and radiation of electromagnetic waves (far fields). For small NPs (< 20 nm), the first channel predominates [1]. For this reason, spasers have never been considered as efficient source of radiation but rather as systems that create high local intensity of the electric field and enhance nonlinear effects. Thus, a boost of energy extraction from spasers is of special interest. In addition, due to the spaser's small size, one cannot design a narrow radiation pattern for the emission of a spaser into free space. According to the antenna theory [10], to achieve a narrow radiation pattern, a wide aperture system built of many spasers, is required. Usually, the phases of emitters in antenna oscillations are specified, but controlling an individual antenna is not a simple task in optics. Ideally, a system of antennas should be self-ordering to create in-phase oscillations. This idea was suggested in Ref. [11] in the framework of a simplified approach in which instead of generation, wave scattering on the lattice of NPs was considered. NPs were assumed to interact with the gain medium described by the negative part of dielectric permittivity. Since the effects of saturation were not taken into account, lasing generation could not be described properly.

In this Letter, we show that the near field interaction of TLSs with neighboring NPs leads to mutual synchronization of spaser oscillations in large 2D arrays of spasers. This mutual synchronization arises due to interaction of quantum subsystem of a spaser with plasmonic particles of the other spasers. The synchronization results in superradiance. Until the array size is smaller than the free space wavelength, the interference of radiated fields is constructive and the radiation intensity power increases as $N^2$ with the number of spasers, $N$. For larger systems the interference becomes destructive and the total radiation power is linear in $N$ while the power of radiation per solid angle perpendicular to the plane grows as $N^2$. The $N^2$- dependence is a consequence of superradiance from a subwavelength array and narrowing of the radiation pattern when the size of the array exceeds the free space wavelength.

As the spaser model, we consider an inversely populated TLS positioned near a nano-hole in a metallic film. The nano-hole with a dipole mode plays the role of the NP. The two-dimensional array of spasers is a perforated metallic film with TLSs positioned near each hole, as shown in Fig. 1. The spacing between holes $\Delta$ is much smaller than the wavelength. We consider a square array with sides $L$ so that the number of spasers in the array is $N = (L/\Delta)^2$. In practice, the quantum system can be electrically pumped via either a *p-n* junction or quantum well parallel to the film. However, discrete TLSs are more convenient for modeling.

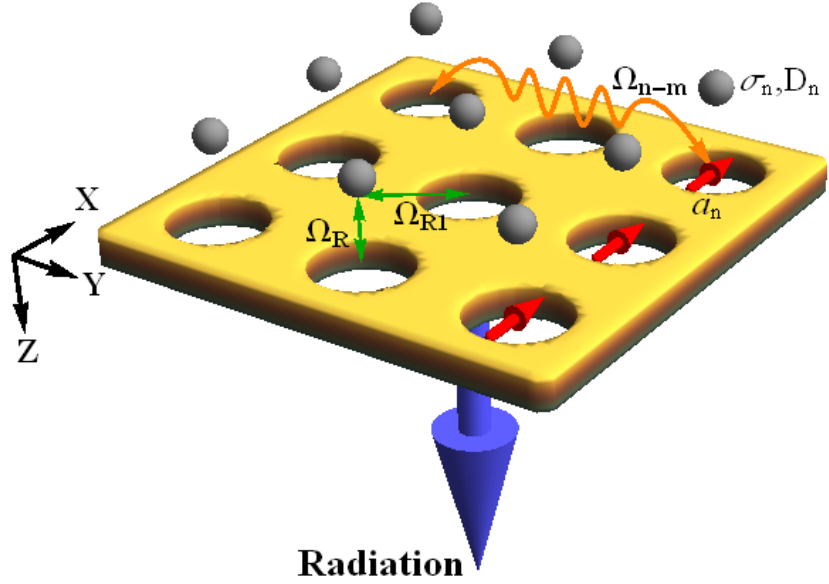

FIG. 1. 2D array of spasers in which quantum dots interact with dipole moments of the holes in the film.

The dynamics of the single spaser is described by a system of three equations for operators of the plasmon amplitude $a_\mathbf{n}$, the TLS polarization $\sigma_\mathbf{n}$ and the population inversion $D_\mathbf{n}$ [12, 13]. As it is shown below, to properly describe radiation of the array, we have to take into account the effect of retardation while considering interaction between the spasers. A spaser located at a point indexed by $\mathbf{n} = \{n_x, n_y\}$ feels the local field $\Omega_{\mathbf{n}-\mathbf{m}} a_\mathbf{m} = \Omega\big((\mathbf{n}-\mathbf{m})\Delta\big) a_\mathbf{m}$ created by a spaser, located at a point $\mathbf{m} = \{m_x, m_y\}$:

$$\Omega(\mathbf{e}R) = \tau_R^{-1} \frac{3}{2}\left( \frac{3(\mathbf{e}\cdot\mathbf{e}_x)^2 - 1}{(k_0 R)^3} - i\frac{3(\mathbf{e}\cdot\mathbf{e}_x)^2 - 1}{(k_0 R)^2} - \frac{(\mathbf{e}\cdot\mathbf{e}_x)^2 - 1}{k_0 R} \right) \exp(ikR), \qquad (1)$$

where $k_0 = \omega/c$ is the wave number of radiation in vacuum, $\mathbf{e}_x$ is the unit vector parallel to the dipole moments (see Fig. 1), $\mathbf{e}R$ is the vector connecting the dipole $\mathbf{m}$ with the dipole $\mathbf{n}$. The local field (1) is the *x*-projection of the electric field created by a unitary dipole parallel to $\mathbf{e}_x$, which is expressed in units of the radiation relaxation rate, $\tau_R^{-1}$. For a spherical nanoparticle of radius $a$ in free space, $\tau_R^{-1} = 2\dfrac{(k_0 a)^3}{\partial \varepsilon / \partial \omega}\bigg|_{\omega=\omega_r}$ [12]. In general, $\tau_R^{-1}$ depends on the environment and the shape on the nanoparticle (or the hole); at the dipole resonance frequency, $\omega_r$, the value of $\tau_R^{-1}$ is related to the polarizability $\alpha(\omega)$ as $\tau_R^{-1} = -\dfrac{2}{3}\dfrac{k_0^3}{\partial \alpha^{-1}/\partial \omega}\bigg|_{\omega=\omega_r}$. The expression for the polarizability of a hole in metal film was obtained in Ref. 14.

In the rotating wave approximation [15-18], the system of interacting spasers is described by the system of equations

$$\dot{a}_\mathbf{n} + \tau_a^{-1} a_\mathbf{n} = -i\Omega_R \sigma_\mathbf{n} - i\Omega_{R1} \sum_{|\mathbf{m}-\mathbf{n}|=1} \sigma_\mathbf{m} + i\sum_{\mathbf{m}\neq\mathbf{n}} \Omega_{\mathbf{n}-\mathbf{m}} a_\mathbf{m}, \qquad (2)$$

$$\dot{\sigma}_\mathbf{n} + \tau_\sigma^{-1} \sigma_\mathbf{n} = i\Omega_R a_\mathbf{n} D_\mathbf{n} + i\Omega_{R1} \sum_{|\mathbf{m}-\mathbf{n}|=1} a_\mathbf{m} D_\mathbf{m}, \qquad (3)$$

$$\dot{D}_\mathbf{n} + (D_\mathbf{n} - D_0)\tau_D^{-1} = 2i\Omega_R(a_\mathbf{n}^* \sigma_\mathbf{n} - \sigma_\mathbf{n}^* a_\mathbf{n}) + 2i\Omega_{R1} \sum_{|\mathbf{m}-\mathbf{n}|=1} (a_\mathbf{m}^* \sigma_\mathbf{n} - a_\mathbf{m} \sigma_\mathbf{n}^*), \qquad (4)$$

where $\tau_\sigma$ and $\tau_D$ denote relaxation times of the polarization (transverse relaxation) and the population inversion (longitudinal relaxation), respectively, $D_0$ describes pumping of a TLS and corresponds to the population inversion in the absence of NPs. In Eqs. (2)-(4), the Rabi frequency $\Omega_R$ characterizes the interaction between the spaser's NP and TLS. In addition to the interaction between plasmonic NPs, the interaction of the NP with TLSs of the neighboring spasers characterized by $\Omega_{R1}$ is taken into account. This interaction synchronizes of the spaser array [19]. To proceed, we neglect quantum correlations and fluctuations and substitute operators by $c$-numbers. The decay rate of the NP plasmonic mode is, determined by the Joule loss in the NP and radiation, $\tau_a^{-1} = \tau_J^{-1} + \tau_R^{-1}$.

The local field (1) is characteristic for free space. The only effect of the metal film that we take into account is attenuation of the wave in space. This is reflected in the appearance of the imaginary part in the wavevector, $k = k' + ik''$ (for calculations we assume that $k' = k_0$ and $k''/k' = 0.2$).

In Fig. 2, phase distributions for small and large arrays are shown. These distributions are obtained by solving Eqs. (2)-(4) numerically for a stationary state. One can see that in both cases the phase distributions are almost uniform. In the large system, a considerable deviation from a uniform distribution only occurs near the boundaries of the large array in the *x*-direction, which is parallel to the direction of oscillations of the dipole moments (Fig. 2b). This is a boundary effect for which the length scale is on the order of the wavelength. Moreover, the phase exhibits weak spatial oscillations along the direction perpendicular to the direction of the dipoles. These oscillations increase in a lossless system. Both of these effects are connected with a nonuniform change of the effective relaxation times of spasers[1] and do not lead to significant changes of the radiation pattern in the spaser array. A detailed consideration of these effects will be published elsewhere.

The uniform phase distribution is the result of spaser synchronization. This solves the problem of extraction of radiation from spasers. Indeed, in the case of synchronized array, interaction (1) forces the spasers to emit radiation. The phenomenon is more evident for a system of a small size, $k_0 R \ll 1$, for which Eq. (1) becomes

$$\Omega(\mathbf{e}R) \approx \tau_R^{-1} \left( \frac{3}{2} \frac{3(\mathbf{e} \cdot \mathbf{e}_x)^2 - 1}{(k_0 R)^3} + i \right) \tag{5}$$

with $\mathrm{Im}\,\Omega(\mathbf{e}R) \approx \tau_R^{-1}$. When all the dipoles oscillate with the same phase and amplitude, the last term in Eq. (2) can be split into two parts:

$$i \sum_{\mathbf{m} \neq \mathbf{n}} \Omega_{\mathbf{n}-\mathbf{m}} a_\mathbf{m} \approx i a_\mathbf{n} \sum_{\mathbf{m} \neq \mathbf{n}} \mathrm{Re}\,\Omega_{\mathbf{n}-\mathbf{m}} - a_\mathbf{n} (N-1) \tau_R^{-1}. \tag{6}$$

---

[1] The variation of relaxation time leads to the variation of the spasing threshold and autonomic frequency of a spaser. In the synchronization regime, to compensate these effects there appear some energy fluxes. In the direction parallel to the dipole moments, the far fields are equal to zero and energy transfer needs a space gradient of the phase [20]. In the transverse direction, the energy transfer is performed by far field that does not need a phase gradient.

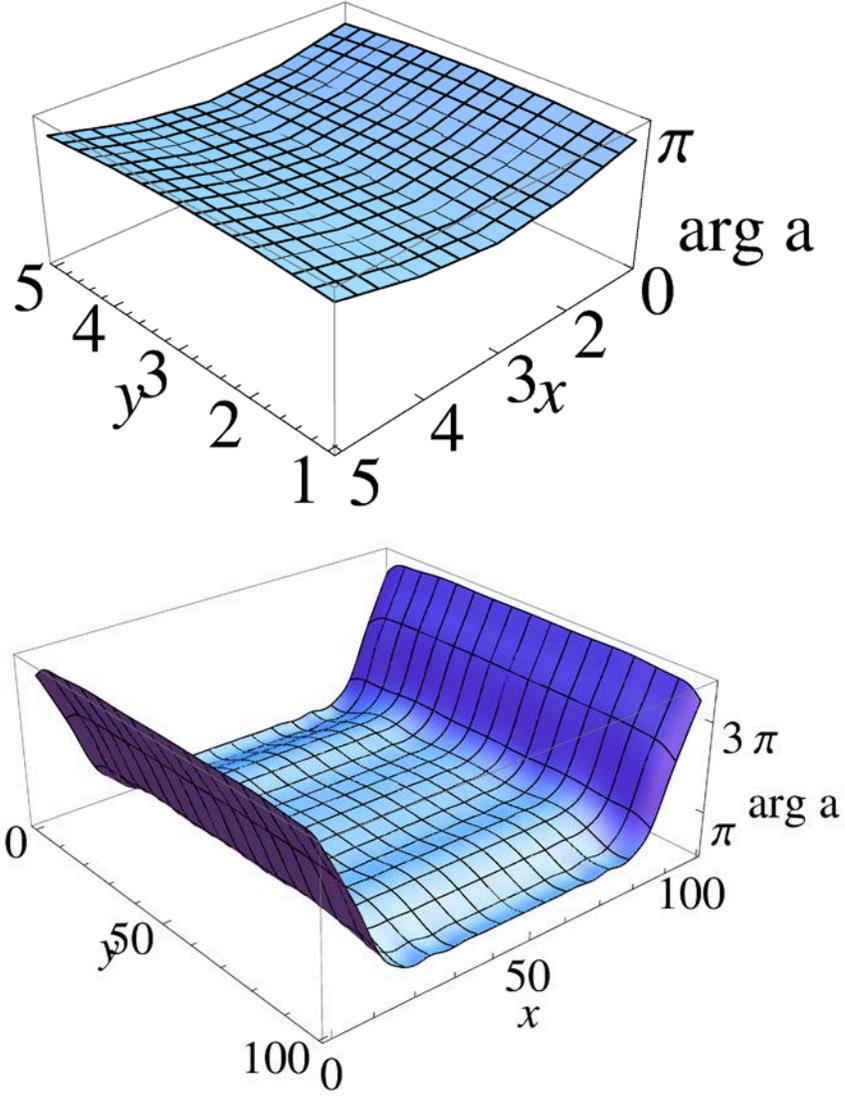

FIG. 2. Phase distribution of the plasmon oscillations in spaser arrays of (a) 5 × 5 and (b) 100 × 100 spasers. In all calculations we use $\Delta = \lambda/20$.

The second term on the right hand side of Eq. (6) leads to an increase of the relaxation rate of the **n**-th plasmon by a factor of $(N-1)\tau_R^{-1}$. As a result, the effective relaxation rate of a plasmon becomes $\tau_J^{-1} + N\tau_R^{-1}$. Equation (2) can now be rewritten as

$$\dot{a}_{\mathbf{n}} + \left(\tau_J^{-1} + N\tau_R^{-1}\right)a_{\mathbf{n}} = -i\Omega_R \sigma_{\mathbf{n}} - i\Omega_{R1}\sum_{|\mathbf{m}-\mathbf{n}|=1}\sigma_{\mathbf{m}} + i\,\mathrm{Re}\sum_{\mathbf{m}\neq\mathbf{n}}\Omega_{\mathbf{n}-\mathbf{m}}a_{\mathbf{m}}. \tag{7}$$

Thus, all plasmonic NPs, in the area with size much smaller than the wavelength contribute equally to the radiation rate giving an effective radiation rate of $\tau_{R\_eff}^{-1} = N\tau_R^{-1}$. The total intensity of radiation of $N$ NPs is then proportional to $N^2$, which is characteristic of superradiance [21-23]. Let us note that to obtain this dependence we use the full dipole field, Eq. (1), including the retardation effects. Thus, all terms in Eq. (1) are required to correctly account for radiative damping in the dipole (spaser) array.

The radiation power of the array can be evaluated by using the energy balance equation, which follows from Eq. (2) for the stationary regime:

$$\sum_{\mathbf{n}}\left[\Omega_R \operatorname{Im}\left(a_\mathbf{n}^* \sigma_\mathbf{n}\right) + \Omega_{R1} \sum_{|\mathbf{m}-\mathbf{n}|=1} \operatorname{Im}\left(a_\mathbf{n}^* \sigma_\mathbf{m}\right)\right] = \tau_J^{-1} \sum_{\mathbf{n}} |a_\mathbf{n}|^2 + \sum_{\mathbf{n},\mathbf{m}} \operatorname{Im}(\Omega_{\mathbf{n}-\mathbf{m}}) \operatorname{Re}(a_\mathbf{n}^* a_\mathbf{m}). \quad (8)$$

Here we assume that at $\mathbf{n}=\mathbf{m}$, $\operatorname{Im}\Omega_{\mathbf{n}-\mathbf{m}} = \tau_R^{-1}$, which corresponds to the limit of short distances in Eq. (5). The left-hand side of Eq. (8) is proportional to the power supplied into the system by TLSs. The first term in the right hand side corresponds to Joule losses. It can be shown that the second term, up to a factor independent of the dipoles' amplitudes, equals the radiation power $I$,

$$I = \frac{m\omega^2}{e^2} \sum_{\mathbf{n},\mathbf{m}} \operatorname{Im}(\Omega_{\mathbf{n}-\mathbf{m}}) \operatorname{Re}(a_\mathbf{n}^* a_\mathbf{m}). \quad (9)$$

This term normalized by the radiation power of a single spaser is shown in Fig. 3a.

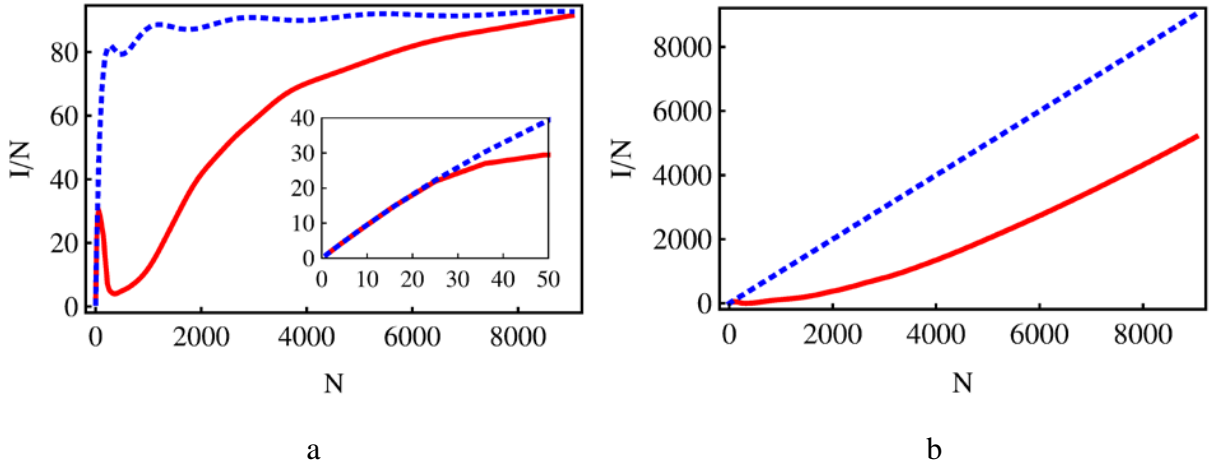

a

b

FIG. 3. Intensity of radiation per spaser for the array with calculated amplitude distribution (solid line) and the array of ideally synchronized dipoles (dashed line) as a function of the array size. (a) The integral intensity. (b) The intensity of radiation normal to the array, $I_\Omega = \partial I / \partial \Omega$. The dependence for small a number of spasers is magnified in the inset in Fig 3a.

As long as the array size is smaller than a half-wavelength ($N < 100$ for our parameters), all spasers are synchronized and oscillate in phase (Fig. 2a). In this case, as one can see in Fig. 3a, the intensity growth is characteristic of superradiance, $I/N \propto N$. Also, for a small array the inset of Fig. 3a shows that radiation from the array nearly coincides with the radiation of a system of synchronized classical dipoles with uniform amplitudes over the array equal to the average amplitude of a spaser $\langle a_\mathbf{n} \rangle$. The radiated intensity drops as the array size increases. The reason for this is a decrease of an effective system size due to boundary effects (see Fig. 2b). An increase of the array size to $L \gg \lambda$ leads to synchronization of spasers for most of the array, as shown in Fig. 2b. Now, radiation per spaser of the array tends to catch up with the radiation of the ideally synchronized system ($N > 5000$ in Fig. 3a). Due to destructive interference of radiation emitted by different parts of a large array, radiations of both spaser and ideally synchronized dipole arrays saturate, so that total radiation becomes proportional to the array size, $I \propto N$.

The important quantity for applications is the intensity per solid angle radiated in the direction normal to the array plane. This is the quantity measured by a detector of a small size. However, calculating energy consumption of the system of dipoles in a local field using Eq. (9) one cannot obtain the anglular distribution of radiation. In order to find this, one should use an electrodynamics analysis of radiation for the obtained distributions of amplitudes and phases for dipole emitters instead of considering the spaser array as a quantum mechanical system. In other words, the spaser array should be considered as an antenna. The distribution of power radiated by the antenna into a solid angle $d\Omega$ in the direction of a unit vector $\mathbf{e}$, $I_\Omega(\mathbf{e})$, is referred to as the radiation pattern and can be determined by the Fourier transform of the field distribution in the antenna opening [10]. In our case, this is the dipole moment distribution in the array. Thus, the radiation pattern is equal to

$$I_\Omega(\mathbf{e}) = I_\Omega^{(0)}(\mathbf{e}) \left| \sum_\mathbf{n} a_\mathbf{n} \exp(-ik_0 \mathbf{e} \cdot \mathbf{r_n}) \right|^2 \tag{10}$$

where $a_\mathbf{n}$ is the stationary solution of Eqs. (2)-(4), $\mathbf{r_n}$ is the position vector of the $\mathbf{n}$-th dipole in the array and $I_\Omega^{(0)}(\mathbf{e}) = (8\pi)^{-1} c k_0^4 \left| [\mathbf{e} \times \mathbf{e}_x] \right|^2$ is the radiation pattern for a unitary dipole. The

integration of Eq. (10) over directions of **e** gives Eq. (9). For the normal direction, $I_\Omega^{(0)}(\mathbf{e})$ is just the sum $I_\Omega = I_\Omega^{(0)}(\mathbf{e}_z)\left|\sum_\mathbf{n} a_\mathbf{n}\right|^2$, which gives $I_\Omega/N \sim N$ for a synchronized array of any size.

This is confirmed by numeric calculations shown in Fig. 3b. Interestingly, the linear increase of $I_\Omega/N$ is determined by two effects. For small arrays, $L < \lambda$, this happens due to the growth of the integral intensity, $I/N$, resulting from superradiance. When $L > \lambda$, $I/N$ saturates and the growth of radiation in the normal direction $I_\Omega$ is caused by the narrowing of the radiation pattern shown in Fig. 4. The latter is due to the growth of the aperture of the radiative system.

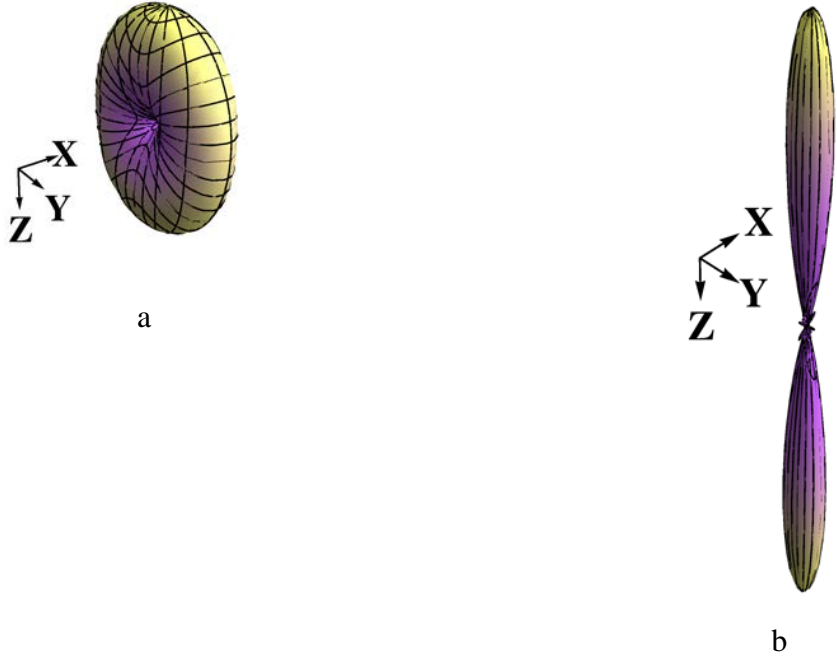

FIG. 4. Radiation pattern for array size of (*a*) 5 × 5 and (*b*) 100 × 100 spasers.

To summarize, a 2D array of spasers is a highly directional source of radiation and allows for an increase of the intensity of radiation by two orders of magnitude compared to a single spaser. Since synchronization occurs over a large area, it is likely that an array of spasers would allow one to create a highly coherent light beam in the transverse direction with a large cross section.

This work was supported by RFBR Grants Nos. 10-02-91750, 11-02-92475 and 12-02-01093 and by a PSC-CUNY grant.